\documentclass[a4paper,12pt]{article}
\usepackage{amssymb,amsfonts}
\usepackage{epsfig}

\begin{document}

\title{Kinks, chains, and loop groups in the $\mathbb{CP}^n$ sigma models}
\author{Derek Harland\footnote{email address: harland@itp.uni-hannover.de}
  \bigskip
  \\Institut f\"ur Theoretische Physik,
  \\Leibniz Universit\"at Hannover,
  \\Appelstra\ss e 2,
  \\30167 Hannover
  \\Germany}
\date{16th October 2009}

\maketitle

\begin{abstract}
We consider topological solitons in the $\mathbb{CP}^n$ sigma models in two space dimensions.  In particular, we study ``kinks'', which are independent of one coordinate up to a rotation of the target space, and ``chains'', which are periodic in one coordinate up to a rotation of the target space.  Kinks and chains both exhibit constituents, similar to monopoles and calorons in $SU(n)$ Yang-Mills-Higgs and Yang-Mills theories.  We examine the constituent structure using Lie algebras.
\end{abstract}

\section{Introduction}

Topological solitons in field theories over $\mathbb{R}^d$ are typically classified by an integer-valued topological charge $N$, and are usually thought of as being some superposition of $N$ basic building blocks, the 1-solitons.  If the space $\mathbb{R}^d$ is replaced by a cylinder $\mathbb{R}^{d-1}\times S^1$ the picture is not so simple: there may be more than one topological charge, while the 1-solitons from the original theory may break up into smaller objects.  The most well-studied case is that of calorons \cite{ll,vb98} in Yang-Mills theory on $\mathbb{R}^3\times S^1$.  While a caloron sometimes resembles a collection of instantons on a cylinder, it is more correctly thought of as a superposition of fundamental monopoles, since instantons may be separated into monopoles, but the converse is not true.  Moreover, there is not just $1$, but $n$ different types of fundamental monopole, $n$ being the rank of the gauge group $SU(n)$.

As a general rule, instantons in $SU(n)$ Yang-Mills theory are similar to the so-called ``lumps'' of the $\mathbb{CP}^{n-1}$ sigma models.  Therefore, it seems likely that lumps on a cylinder $\mathbb{R}\times S^1$ (or ``chains'') will have similar properties to calorons.  In a recent paper \cite{bruckmann08}, Bruckmann has constructed an example of a chain in the $\mathbb{CP}^1$ sigma model which exhibits constituents in much the same way as calorons.  Here we extend Bruckmann's analysis to $\mathbb{CP}^n$ sigma models.  Further aspects of $\mathbb{CP}^n$ chains are explored in another new article \cite{bruckmann09}.  After completing this work, we learned that similar constructions also appeared in the earlier paper \cite{efinoos06}.

It is worth asking why constituents were not discovered in sigma models earlier.  In fact, there has been a number of studies of chains in sigma models \cite{mottola&wipf89,snippe94,romao05}.  The reason constituents were not observed is that the constituents are only visible if the periodic boundary condition is accompanied by a suitable ``twist''; this is entirely analogous to the condition of ``non-trivial holonomy'' which is necessary for constituent monopoles to be observed in calorons.

A number of explanations have been put forward for the constituent structure of calorons.  The earliest can be found in the work of Garland and Murray, who, following a suggestion of Hitchin, showed that a caloron can be identified with a monopole on $\mathbb{R}^3$ whose gauge group is a loop group \cite{gm88}.  Then the different types of constituent in a caloron can be associated with the roots of the loop group, just as an $SU(n)$ monopole exhibits constituents associated with the roots of $SU(n)$ \cite{weinberg80}.  Unaware of this earlier work, Lee and Yi later predicted the existence of constituents using a string-theoretic argument \cite{lee&yi}, and this paper motivated the explicit constructions of \cite{ll,vb98}.  The Nahm transform for calorons is also suggestive of a constituent structure, as noted by Kraan and van Baal \cite{vb98}.  However, the link between constituents and Nahm transforms is not universal: monopole chains have both constituents and a Nahm transform \cite{ward05}, but there is no clear link between the two.

Prior to the explicit constructions \cite{bruckmann08,efinoos06}, chains in the $\mathbb{CP}^n$ sigma models were predicted to exhibit constituents by Tong \cite{tong02}, using string theory arguments similar to Lee and Yi's.  Tong's argument actually applied to semi-local vortices, which approach sigma model lumps in a strong coupling limit, and motivated the paper \cite{efinoos06}.  Here, we will show that, like the constituent monopoles of a caloron, the constituents of a sigma model chain can also be understood classically, using loop groups.  Unlike in calorons, consituents in sigma model chains cannot be understood using a Nahm transform, since no Nahm transform for sigma models is known \cite{aaw02}.

An outline of the rest of this article is as follows.  In section 2, we briefly recall the definition of the $\mathbb{CP}^n$ sigma models.  In section 3, we review and analyse the sigma-model analog of monopoles, which we have called ``multi-kinks'', and which were previously studied in \cite{gtt01}.  In section 4 we construct and analyse chains in the sigma models.  In section 5 we review some facts about loop groups, and in section 6 we show how these can be used to understand the constituent structure of chains.  We conclude with some comments in section 7.

\section{The $\mathbb{CP}^n$ sigma models}

The field content of a sigma model is a map $\phi:M\rightarrow N$ between two manifolds equipped with metrics $g_M$, $g_N$, and induced volume forms $V_M$, $V_N$.  The energy functional is
\begin{equation} E = \frac{1}{2}\int_{M} \| d\phi \|^2 V_M. \end{equation}
Here, and throughout, we use the shorthand notation 
\begin{equation}
\|d\phi\|^2=(g_M)^{ij}(g_N)_{ab} \frac{\partial\phi^a}{\partial x^i} \frac{\partial\phi^b}{\partial x^j}.
\end{equation}
In the mathematical literature, a function $\phi$ which minimizes $E$ is called harmonic.  In the case where $M$ is a Riemannian 2-manifold and $N$ is K\"{a}hler with K\"ahler form $\omega_N$, there is a lower bound
\begin{equation} E \geq \int_{M} \phi^\ast(\omega_N), \end{equation}
called the Bogomolny bound, which is saturated if and only if $\phi$ is holomorphic \cite{guest85}.  The lower bound is homotopy-invariant.

We will be interested in the case where $M$ is $\mathbb{C}$ and $N$ is $n$-dimensional complex projective space $\mathbb{CP}^n$.  The simplest way to define this space is as a quotient: $\mathbb{CP}^n=SU(n+1)/U(n)$.  In practise, we will use two models for this space.  The first is as an adjoint orbit: consider the adjoint action of $SU(n+1)$ on the Lie algebra $su(n+1)$ of traceless anti-Hermitian matrices.  The centraliser of the element $b=-i/(n+1)^2\, diag (n,-1,\dots,-1) \in su(n+1)$ is $U(n)$, therefore $\mathbb{CP}^n$ can be identified with the orbit of $b$ under the action of $SU(n+1)$.  A natural metric on $su(n+1)$ is given by the Cartan-Killing form, denoted $\langle\cdot,\cdot\rangle$ \footnote{Note that $\langle Y,Z \rangle = -2(n+1)Tr(YZ)$.}.  The metric $g$ on the adjoint orbit is chosen proportional to that obtained by restriction:
\begin{equation}
\label{metric}
g_p(X,Y) := (n+1)\langle X,Y\rangle \quad \forall p \in \mathbb{CP}^n, X,Y \in T\mathbb{CP}^n.
\end{equation}

A second model for $\mathbb{CP}^n$ is as the quotient of $\mathbb{C}^{n+1}\backslash\{0\}$ by the action of $\mathbb{C}\backslash\{0\}$, where $\mathbb{C}\backslash\{0\}$ acts on vectors by multiplication.  In this model we represent points in $\mathbb{CP}^n$ by non-zero column vectors with square brackets, and column vectors related by a rescaling are understood to be equivalent:
\begin{equation}
\label{rescaling}
\left[ \begin{array}{c} z^1 \\ \vdots \\ z^n \end{array} \right] \sim
\left[ \begin{array}{c} \kappa z^1 \\ \vdots \\ \kappa z^n \end{array} \right] ,\quad \kappa\in\mathbb{C}\backslash \{0\}.
\end{equation}
This representation makes it easy to write down holomorphic functions.  For example, the charge 1 $\mathbb{CP}^1$ sigma model lump may be written
\begin{equation}
\label{CP1 lump}
\phi: u \mapsto \left[ \begin{array}{c} \lambda \\ u-a \end{array} \right] 
\end{equation}
where $u=x+iy$ is a coordinate on $\mathbb{C}$ and $\lambda \in \mathbb{C}^\ast$, $a\in\mathbb{C}$ are parameters.  For more details, see \cite{manton&sutcliffe}.

A disadvantage of this representation is that it can be tricky to evaluate the energy density of a given field $\phi$.  One must first use the rescaling (\ref{rescaling}) to ensure $\phi$ takes values in $S^{2n+1}$, and then evaluate the $\| d\phi \|^2$ using the quotient metric on $S^{2n+1}/U(1)$.  The quotient metric agrees with (\ref{metric}) when $S^{2n+1}$ is defined by $\sum |z^j|^2 = 8n/(n+1)$.  Thankfully, we will rarely need to evaluate energy densities in what follows.

\section{Multi-kinks}

Multi-kinks were introduced in \cite{gtt01}, and are the sigma model analog of monopoles.  We define a multi-kink to be a function $\tilde{\phi}:\mathbb{R}\rightarrow \mathbb{CP}^n$ which minimizes the energy functional,
\begin{equation}
\label{mdw energy 2}
E = \frac{1}{2}\int_\mathbb{R} \| \partial_x\tilde{\phi} \|^2 + V(\tilde{\phi}) dx.
\end{equation}
The potential function $V\colon\mathbb{CP}^n\rightarrow\mathbb{R}$ is determined by a Lie algebra element $X\in \mathfrak{su}(n+1)$ as follows.  The action of $X$ induces a vector field $\tilde{X}$ on $\mathbb{CP}^n$.  The value of $V$ at a point $z\in\mathbb{CP}^n$ is equal to the length squared of $\tilde{X}$ at $z$: $V(z)=\|\tilde{X}(z)\|^2$.  We will evaluate $V$ explicitly below.

Given a multi-kink $\tilde{\phi}$, we define $\phi(x,y)=\exp(-yX)\tilde{\phi}(x)$.  In terms of $\phi$, the energy functional takes the simpler form
\begin{equation}
\label{mdw energy 1}
E = \frac{1}{2}\int_\mathbb{R} \| d\phi \|^2 dx.
\end{equation}
Although $\phi$ depends on both $x$ and $y$, $\phi$ should still be regarded a 1-dimensional object, because the dependence on $y$ is trivial:
\begin{equation} \frac{\partial\phi}{\partial y} = -X\cdot\phi \end{equation}
Notice that, since $\phi$ winds in the spatial $y$-direction, multi-kinks are rather similar to Q-kinks \cite{abraham&townsend1,abraham&townsend2}, which wind in the time direction.

We may choose coordinates so that $X = i\,diag(\mu_1,\dots \mu_{n+1})$ for real numbers $\mu_i$ satisfying $\sum\mu_i=0$ and $\mu_i\geq\mu_{i+1}$.  We assume further that these inequalities are strict: this is analogous to the condition of maximal symmetry breaking for monopoles.  In order that the energy be finite, $\phi$ must tend to a fixed point of $X$, called a vacuum, as $x\rightarrow\pm\infty$.  It is easy to see that there are $n+1$ vacua in $\mathbb{CP}^n$, written
\begin{equation} v_1 = \left[ \begin{array}{c} 1 \\ 0 \\ \vdots \\ 0  \end{array} \right],\,
v_2 = \left[ \begin{array}{c} 0 \\ 1 \\ \vdots \\ 0  \end{array} \right], \cdots ,
v_{n+1} = \left[ \begin{array}{c} 0 \\ 0 \\ \vdots \\ 1  \end{array} \right]. \end{equation}
Although not needed later, we mention here that the potential function $V$ takes the form,
\begin{equation}
V = \frac{8n}{n+1} \frac{ \left(\sum_j |z^j|^2\right) \left(\sum_j \mu_j^2|z^j|^2\right) - \left(\sum_j \mu_j |z^j|^2\right)^2 }{ \left( \sum_j |z^j|^2 \right)^2 },
\end{equation}
which is invariant under rescalings of $z^j$ (\ref{rescaling}), as it should be.  The factor $8n/(n+1)$ is a consequence of our choice of metric (\ref{metric}).  The vacua $v_j$ are of course zeros of $V$.

There is a Bogomolny lower bound on the energy,
\begin{equation}
\label{dw bog bound}
E \geq \int_{\mathbb{R}} \phi^{\ast} i_{\tilde{X}} \omega, 
\end{equation}
whose proof will be deferred to section 4.  Here $\omega$ is the standard K\"{a}hler form on $\mathbb{CP}^n$ and $i_{\tilde{X}}\omega$ its inner derivative, defined by $i_{\tilde{X}}\omega(Y)=\omega(\tilde{X},Y)$.  The bound is saturated if and only if $\phi$ is holomorphic,
\begin{equation}
\label{dw bog equation 2}
\frac{\partial}{\partial x} + i \frac{\partial}{\partial y} \phi = 0,
\end{equation}
or equivalently $\tilde{\phi}$ solves the Bogomolny equation,
\begin{equation}
\label{dw bog equation}
\frac{\partial}{\partial x} \tilde{\phi} = iX\cdot\tilde{\phi}.
\end{equation}
Since the vector field $X$ is Hamiltonian, there locally exists a Hamiltonian function $\psi$ on $\mathbb{CP}^n$ such that $i_{\tilde{X}}\omega=d\psi$.  Actually, this function $\psi$ is globally well-defined, and can be constructed explicitly.

To construct the function $\psi$, we work in the adjoint orbit model of $\mathbb{CP}^n$.  Let $su(n+1)$ denote the Lie algebra of traceless anti-Hermitian matrices, and let $\mathfrak{t}$ denote the Cartan subalgebra of diagonal matrices.  The point $b = -i/(n+1)^2\, diag (n,-1,\dots,-1) \in \mathfrak{t}$ has stabiliser $U(n)$ under the adjoint action of $SU(n+1)$, so its orbit may be identified with $\mathbb{CP}^n$ ($b$ itself is identified with $v_1$).

The function $\psi$ is defined in terms of the Cartan-Killing form $\langle \cdot , \cdot\rangle$ by
\begin{equation} \psi(p) = -\langle p, X \rangle \end{equation}
for $p$ in the orbit of $b$.  The canonical choice of symplectic form (due to Kostant and Kirillov) is
\begin{equation} \omega_p(\tilde{\xi}_p,\tilde{\eta}_p) = -\langle p, [\xi,\eta] \rangle, \end{equation}
where $\tilde{\xi}_p=[\xi,p]$ and $\tilde{\eta}_p=[\eta,p]$ are the tangent vectors at $p$ induced by the action of $\xi,\eta\in su(n+1)$.  This $\omega$ is compatible with the metric (\ref{metric}) in the usual sense.  The identity $i_{\tilde{X}}\omega=d\psi$ is equivalent to
\begin{equation}
\label{hamiltonian}
\omega_p(\tilde{X}_p,\tilde{Y}_p) = L_{\tilde{Y}_p} \psi \quad \forall Y\in su(n+1)
\end{equation}
where $L$ denotes the Lie derivative.  We have $L_{\tilde{Y}_p}\psi = -\langle [Y,p],X \rangle$ and $\omega_p(\tilde{X}_p,\tilde{Y}_p) = -\langle p,[X,Y] \rangle$, so the identity (\ref{hamiltonian}) follows from the invariance of the Killing form:
\begin{equation} \langle [Y,p],X \rangle + \langle p, [Y,X]\rangle = 0. \end{equation}

So, for a multi-kink $\phi$ satisfying the boundary conditions $\phi\rightarrow v_{j_\pm}$ as $x\rightarrow\pm\infty$, the lower bound (\ref{dw bog bound}) is
\begin{equation} \int_{\mathbb{R}} \phi^{\ast} i_{\tilde{X}} \omega = \psi(v_{j_+}) - \psi(v_{j_-}) = 2(\mu_{j_-}-\mu_{j_+}). \end{equation}
where we have used the fact that $\psi(v_j)=-2\mu_j$.

The simplest case of the above is when $n=1$.  Since $X$ must be traceless, $\mu_1>0$ and $\mu_2=-\mu_1$.  The general solution of the Bogomolny equation (\ref{dw bog equation 2}) is
\begin{equation}
\label{fundamental wall}
\phi: u \mapsto \left[ \begin{array}{c} a_1\exp(-\mu_1u) \\ a_2\exp(-\mu_2u) \end{array} \right]
\end{equation}
where $u=x+iy$ and $a_1,a_2\in\mathbb{C}^\ast$ are defined up to rescaling.  In order to evaluate the limits of this solution at $x=\pm\infty$, it should be appropriately rescaled using (\ref{rescaling}): it turns out that $\phi \rightarrow v_1$ as $x\rightarrow-\infty$ and $\phi \rightarrow v_2$ as $x\rightarrow\infty$.  We define the mass $\nu$ of the kink to be half its energy: $\nu:=\mu_1-\mu_2$.

Actually, this $\mathbb{CP}^1$ kink is the familiar sine-Gordon kink.  Consider a kink of the form,
\begin{equation} \tilde{\phi}: x\mapsto \left[ \begin{array}{c} 2\cos(f(x)/2) \\ 2 e^{i\alpha(x)} \sin(f(x)/2) \end{array} \right], \end{equation}
for a real functions and $f$ and $\alpha$.  The energy (\ref{mdw energy 2}) of such a kink is
\begin{equation} E = \frac{1}{2} \int_{\mathbb{R}} (\partial_x f)^2 + (\nu^2+(\partial_x \alpha)^2) \sin^2(f) dx. \end{equation}
Clearly this energy is minimized when $\partial_x \alpha=0$, and after substituting this equation the energy functional of the sine-Gordon model is recovered.  In particular, the kink (\ref{fundamental wall}) corresponds to the sine-Gordon kink.  We define the location of the kink to be the point $x=x_0$ where $f(x)=\pi/2$, that is, $x_0=\nu^{-1}\ln|a_1/a_2|$.  The second modulus of the kink corresponds to a $U(1)$ phase.

The next simplest case is $n=2$.  The general solution of the Bogomolny equation is
\begin{equation}
\phi: u \mapsto \left[ \begin{array}{c} a_1\exp(-\mu_1u) \\ a_2\exp(-\mu_2u) \\ a_3\exp(-\mu_3u) \end{array} \right]
\end{equation}
where $a_1,a_2,a_3\in\mathbb{C}$ are defined up to rescaling, and at least two are non-zero.  More concisely, we can say that $[a_1,a_2,a_3]^t \in \mathbb{CP}^2\setminus V$, where $V=\{v_1,v_2,v_3\}$ is the set of vacua.  When $a_3=0$, the solution satisfies the boundary condition $\phi(x=-\infty)=v_1$, $\phi(x=\infty)=v_2$, and the solution is an embedding of a $\mathbb{CP}^1$ kink with mass $\nu_1:=\mu_1-\mu_2$.  Similarly, when $a_1=0$ the solution is an embedding of the $\mathbb{CP}^1$ kink satisfying the boundary conditions $\phi(x=-\infty) = v_2$, $\phi(x=\infty)=v_3$ and with mass $\nu_2:=\mu_2-\mu_3$.  In all other cases, the boundary condition satisfied is $\phi(x=-\infty) = v_1$ and $\phi(x=\infty)=v_3$, and the mass of the solution is $\nu_1+\nu_2$.

The kinks with with masses $\nu_1$ and $\nu_2$ can be obtained as limits of the multi-kink with mass $\nu_1+\nu_2$.  They are ``fundamental'', in the sense that they cannot be decomposed further into smaller kinks.  The 4 moduli of the general solution can be accounted for by assigning 2 to each constituent fundamental kink.  So it makes sense to think of the multi-kink as a superposition of two fundamental kinks.  For suitable values of the parameters, one can identify the two constituent kinks as isolated lumps of energy density.  The mass $\nu_1$ kink will always lie to the left of the mass $\nu_2$ kink on the $x$-axis, since the the former tunnels from $v_1$ to $v_2$ while the latter tunnels from $v_2$ to $v_3$.

It is easy to see how to generalise the $n=2$ multi-kink, and its fundamental kink constituents, the cases $n>2$.  The general picture is that a $\mathbb{CP}^n$ multi-kink is a superposition of $n$ fundamental kinks of masses $\nu_i:=\mu_i-\mu_{i+1}$, and possesses $2n$ moduli.  There is still an ordering prescription: the mass $\nu_i$ kink lies to the left of the mass $\nu_{i+1}$ kink.

The structure of multi-kinks can be visualised using the adjoint orbit model of $\mathbb{CP}^n$.  The element $X\in \mathfrak{t}$ determines a fundamental Weyl chamber (the chamber containing $X$), and hence sets of positive and simple roots.  The simple roots are $\alpha_i : diag(a_1,\dots,a_{n+1}) \mapsto a_i-a_{i+1}$.  The point $b$, which was identified with $v_1$, lies on the boundary of the fundamental Weyl chamber.  The other vacua $v_j$ are identified with the images of $b$ under the action of the Weyl group.

Any root $\alpha$ determines an embedding of $su(2)$ in $su(n+1)$, and hence a homorphism from $SU(2)$ into $SU(n+1)$.  If $v_j\in\mathfrak{t}$ is a vacuum which is not fixed by this $SU(2)$, then its orbit under the action of $SU(2)$ may be identified with $\mathbb{CP}^1$.  Thus a root and a well-chosen vacuum determine a map from $\mathbb{CP}^1$ to $\mathbb{CP}^n$, which is in fact holomorphic, and a $\mathbb{CP}^n$ kink can be obtained by composing this map with a $\mathbb{CP}^1$ kink.  The embedded kink will interpolate between $v_j$ and its image under the element $s_\alpha$ of the Weyl group associated with $\alpha$, and its mass will be
\begin{eqnarray}
\psi(s_\alpha(v_j))-\psi(v_j) &=& -\langle X, s_\alpha(v_j)-v_j \rangle \\
\label{mass formula}
&=& \frac{2\alpha(X)\alpha(v_j)}{\left\langle \alpha,\alpha \right\rangle }
\end{eqnarray}
In particular, the fundamental kinks described above are determined by the simple roots of $su(n+1)$.  The indecomposability of a fundamental kink is linked with the indecomposability of a simple root.

\begin{figure}[ht]
\epsfig{file = 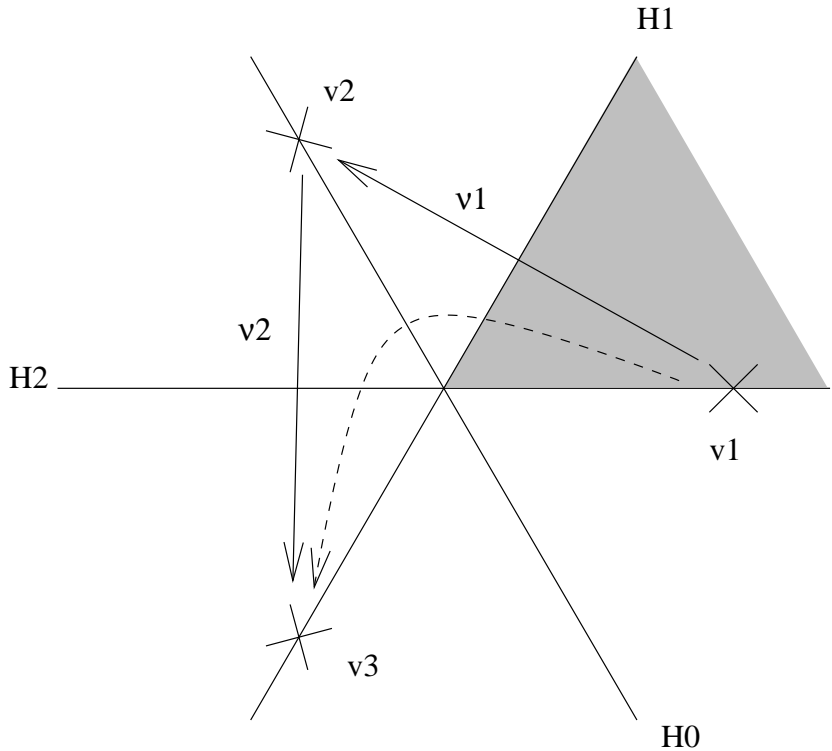, scale = 1}
\caption{The $\mathbb{CP}^2$ multi-kink realised as an adjoint orbit in the Lie algebra of $SU(3)$.}
\label{smc fig 1}
\end{figure}

We have illustrated this for the $\mathbb{CP}^2$ case in figure \ref{smc fig 1}: the Cartan subalgebra $\mathfrak{t}$ is two-dimensional, and is represented by the plane.  The shaded region represents the fundamental Weyl chamber, and the point $v_1$ is indicated by a cross.  The Weyl group is generated by reflections in the lines $H_1$, $H_2$, $H_0$.  The other vacua $v_2$, $v_3$ are the images of $v_1$ under the action of the Weyl group.  The multi-kink, which interpolates between $v_1$ and $v_3$, is indicated by a dashed arrow, while the fundamental kinks are indicated by the solid arrows.

\section{Chains}

A chain is a map $\phi: \mathbb{R}^2 \rightarrow \mathbb{CP}^n$ satisfying the periodicity condition
\begin{equation}
\label{periodicity condition}
\phi(x,y+\beta) = g \cdot \phi(x,y),
\end{equation}
which minimizes the energy functional,
\begin{equation} E = \frac{1}{2} \int_0^\beta \int_{\mathbb{R}} \| d\phi \|^2 dx\,dy. \end{equation}
Here $\beta>0$ is the period and $g\in SU(n+1)$.  In order that the energy be finite, $\phi$ must tend to fixed points of $g$ as $x\rightarrow\pm\infty$.  The periodicity condition (\ref{periodicity condition}) is chosen to mimic a similar condition satisfied by calorons in the ``algebraic gauge'' \cite{vb98} (where the value of the component of the gauge field in the periodic direction is made to vanish at infinity).

We assume that $g$ is written in the form $g=\exp(-X\beta)$, with $X=idiag(\mu_1,\dots \mu_{n+1})$.  We assume that $\mu_i > \mu_{i+1}$ and $\mu_1-\mu_{n+1} < \mu_0$, where $\mu_0:=2\pi/\beta$.  We do not consider at present the possibility that some of the $\mu_i$ are equal.  Then the fixed points of $g$ are the vacua $v_1,\dots, v_{n+1}$ described in the preceding section.

As usual, there is a Bogomolny bound on the energy,
\begin{equation} E \geq \int_{\mathbb{R}\times S^1} \phi^\ast(\omega), \end{equation}
and solutions of the Bogomolny equation are holomorphic.  The lower bound is evaluated as follows.  Let $\tilde{\phi}(x,y)=\exp(yX)\phi(x,y)$; then $\tilde{\phi}$ is strictly periodic with well-defined limits as $x\rightarrow \pm\infty$, hence extends to a map from $S^2$ to $\mathbb{CP}^n$.  Let $k_0\in\mathbb{Z}$ denote the degree of this map.  Using the chain rule, and the fact that the action of $X$ fixes the symplectic form $\omega$, we obtain an identity
\begin{equation} \tilde{\phi}^\ast \omega  = \phi^\ast \omega  - \phi^*(i_{\tilde{X}} \omega) \wedge dy. \end{equation}
We recall from earlier the that $i_{\tilde{X}} \omega = d\psi$.  Integrating, we obtain
\begin{eqnarray}
\int_{\mathbb{R}\times S^1} \phi^\ast(\omega) &=& \int_{\mathbb{R}\times S^1} \tilde{\phi}^\ast(\omega) + \beta\int_\mathbb{R} \phi^\ast(d\psi) \\
&=& 4\pi k_0 + \beta\int_{v_{j_-}}^{v_{j_+}} d\psi \\
\label{caloron charge}
&=& 4\pi k_0 + \frac{4\pi}{\mu_0} (\mu_{j_-}-\mu_{j_+}).
\end{eqnarray}
This formula bears a striking resemblance with the formula for the charge of a caloron: see for example \cite{nye}.  The Bogomolny bound (\ref{dw bog bound}) for multi-kinks is a special case, since multi-kinks satisfy (\ref{periodicity condition}).

A general $\mathbb{CP}^n$ chain may be written
\begin{equation}
\label{general chain}
\phi(u) = \left[ \begin{array}{c} \exp(-\mu_1 u) f_1(\exp(\mu_0 u)) \\ \vdots \\ \exp(-\mu_{n+1}u) f_{n+1}(\exp(\mu_0 u)) \end{array} \right]
\end{equation}
with $f_j$ meromorphic functions.  The $f_j$ can be chosen rational, since if this is not the case the chain will have infinite energy.  By multiplying through by denominators, we can write the non-zero $f_j$ as polynomials,
\begin{equation} f_j(w) = \sum_{k=m_j}^{n_j} a_{jk} w^k, \end{equation}
for complex numbers $a_{jk}$ and non-negative integers $m_j,n_j$, such that $a_{jm_j},a_{jn_j}\neq0$, $\min\{m_j|f_j\neq0\}=0$, and the non-zero polynomials $f_j$ have no common root.  This representation of a general chain is unique up to overall rescaling of the coefficients $a_{jk}$.

The general chain (\ref{general chain}) satisfies the boundary condition $\phi\rightarrow v_{j_\pm}$ as $x\rightarrow\pm\infty$, where $j_+ = \max\{j |f_j\neq0, n_j=\max\{n_k\}\}$ and $j_- = \min\{j |f_j\neq0, m_j=0\}$.  The charge $k_0$ is equal to $\max\{n_j|f_j\neq0\}$.  To see this, recall that $k_0$ is defined to be the degree of the map
\begin{equation} u \mapsto \left[ \begin{array}{c} \exp(-\mu_1 x) f_1(\exp(\mu_0 u)) \\ \vdots \\ \exp(-\mu_{n+1}x) f_{n+1}(\exp(\mu_0 u)) \end{array} \right]. \end{equation}
This map is homotopic to the map,
\begin{equation}
u \mapsto \left[ \begin{array}{c} f_1(\exp(\mu_0 u)) \\ \vdots \\ f_{n+1}(\exp(\mu_0 u)) \end{array} \right],
\end{equation}
whose degree is $\max\{n_j|f_j\neq0\}$.

From the above parametrisation, we see that a $\mathbb{CP}^n$ chain with charge $k_0$ satisfying the boundary conditions $\phi\rightarrow v_{j_\pm}$ as $x\rightarrow\pm\infty$ is determined by $(n+1)k_0 + (j_+-j_-)$ complex parameters.

The simplest examples of chains are the multi-kinks from the previous section.  These are chains with $k_0=0$, and in fact all such chains are multi-kinks.  Thus chains with $k_0=0$ have trivial dependence on $y$; to have non-trivial dependence on $y$, one must have $k_0\neq0$.

Bruckmann's example \cite{bruckmann08} depends non-trivially on $y$.  This example was constructed in the $\mathbb{CP}^1$ model, and has charges $k_0=1$, $j_+=j_-=1$.  It can be parametrised as follows:
\begin{equation}
\label{bruckmann chain}
\phi: u \mapsto \left[ \begin{array}{c} a_{11} \exp((\mu_0-\mu_1)u) + a_{10} \exp(-\mu_1u) \\ a_{20} \exp(-\mu_2u) \end{array} \right],
\end{equation}
for $a_{11}$, $a_{10}$, $a_{20}$ non-zero complex parameters, defined up to an overall scaling.  Note that $\mu_2=-\mu_1$ because $X$ is traceless.

Bruckmann observed \cite{bruckmann08} that, for certain values of the parameters the chain resembles a superposition of two fundamental kinks, with masses $\nu_1=\mu_1-\mu_2$ and $\nu_2=\mu_0-\nu_1$.  To see this, we fix some of the parameters: by making translations and phase rotations, we can choose $a_{11}=-a_{10}$ and we define $b=a_{20}/a_{11}$.  The chain (\ref{bruckmann chain}) becomes
\begin{equation} \phi : u \mapsto \left[ \begin{array}{c}  \exp(\nu_2u)-\exp(-\nu_1u) \\ b \end{array} \right]. \end{equation}
When $|b|$ is small, the field resembles a lump of the form (\ref{CP1 lump}), with location $a=0$ and parameter $\lambda=b/\mu_0$.  When $b$ is large, the field resembles two kinks of the form (\ref{fundamental wall}), with masses $\nu_1,\nu_2$ and locations $x_1=-\nu_1^{-1}\ln |b|$, $x_2 = \nu_2^{-1}\ln |b|$.

The simplest example in which both terms in the formula (\ref{caloron charge}) contribute to the topological charge occurs in the $\mathbb{CP}^1$ sigma model.  It has topological charges $k_0=1$, $j_+=2$, $j_-=1$, and is written
\begin{equation}
\phi: u \mapsto \left[ \begin{array}{c} a_{11} \exp((\mu_0-\mu_1)u) + a_{10} \exp(-\mu_1u) \\ a_{21} \exp((\mu_0-\mu_2)u) + a_{20} \exp(-\mu_2u) \end{array} \right],
\end{equation}
with $a_{10},a_{21}\neq0$.  This chain depends non-trivially on $y$.  For certain choices of the parameters $a_{jk}$, it resembles three kinks, two with mass $\nu_1$ and one with mass $\nu_2$.  The mass-$\nu_2$ kink always lies between the two mass-$\nu_1$ kinks on the $x$-axis.

More generally, a $\mathbb{CP}^n$ chain resembles a superposition of $n+1$ types of fundamental kink.  There are $n$ types of kink with masses $\nu_i=\mu_i-\mu_{i+1}$ for $i=1,\dots,n$ which tunnel from $v_i$ to $v_{i+1}$, and the $(n+1)$th type of kink has mass $\nu_{n+1}= \mu_0-\mu_1+\mu_{n+1}$ and tunnels from $v_{n+1}$ to $v_1$.  If $j_+=j_-$ there are $k_0$ kinks of each type.  If $j_+>j_-$, there are $k_0+1$ kinks of mass $\nu_j$ for $j_- \leq j < j_+$ and $k_0$ kinks of the remaining types.  If $j_+<j_-$, there are $k_0-1$ kinks of mass $\nu_j$ for $j_+ \leq j < j_-$ and $k_0$ kinks of the remaining types.

The moduli of the chain are accounted for by assigning two real moduli to each kink, one for translation and one for an internal ``phase''.  Similarly, the energy of the chain is equal the sum of the masses of the constituent kinks, times a factor $4\pi/\mu_0$.

We noted above that multi-kinks with trivial $y$-dependence are examples of chains.  In the constituent picture, these are chains with no constituents of the $(n+1)$th type.  So the $(n+1)$th kink is necessary to have non-trivial dependence on $y$.

Similar to multi-kinks, there is an ordering prescription among the constituents of a chain.  A kink of mass $\nu_{i+1}$ always lies to the right of a kink of mass $\nu_i$, and a kink of mass $\nu_1$ lies to the right of a kink of mass $\nu_{n+1}$.  A similar ordering prescription was noticed for calorons with $SO(2)$ symmetry: see \cite{vb02,symcal}.  If there are no kinks present with mass $\nu_{n+1}$, the chain is actually a multi-kink with trivial $y$-dependence.

\section{Loop groups}

A loop group $LG$ is a group whose elements are smooth maps from the circle $S^1$ to a Lie group $G$.  The product of two loops is obtained by pointwise multiplication, using the product in $G$.  The standard reference on loop groups is the book by Pressley and Segal \cite{ps86}.  Although loop groups are infinite-dimensional, their root structure is similar to that of finite-dimensional Lie groups, which is part of the reason why they are interesting to study.

We denote by $\mathbb{T}$ the group of rigid rotations of $S^1$.  This acts in a natural way on $LG$ and it is standard practise to define a semi-direct product $\mathbb{T}\tilde{\times}LG$.  The group $\mathbb{T}\tilde{\times}LSU(n+1)$ acts naturally on the space $L\mathbb{CP}^n$ of smooth maps from $S^1$ to $\mathbb{CP}^n$.  In fact, this action is transitive: let us represent an element of $L\mathbb{CP}^n$ by a function $f:[0,2\pi]\rightarrow \mathbb{CP}^n$ satisfying $f(2\pi)=f(0)$.  Since $SU(n+1)$ acts transitively on $\mathbb{CP}^n$, we can choose $g:[0,2\pi]\rightarrow SU(n+1)$ such that $f(\theta)=g(\theta)v_1$.  At this stage $g(\theta)$ need not be periodic, but certainly $g(2\pi)^{-1}g(0)$ is an element of the stabiliser group $U(n)\subset SU(n+1)$ of $v_1$.  Since $U(n)$ is path connected, we can choose $h:[0,2\pi]\rightarrow U(n)$ such that $h(0)$ is the identity and $h(2\pi)=g(2\pi)^{-1}g(0)$.  Then $g'(\theta):=g(\theta)h(\theta)$ defines a continuous map from $S^1$ to $SU(n+1)$ such that $f=gv_1$.  If $f$ is smooth, then $g'$ may also be chosen smooth.  Therefore $LSU(n+1)$ and $\mathbb{T}\tilde{\times}LSU(n+1)$ act transitively on $L\mathbb{CP}^n$.

The stabilisers of the constant loop $\theta\mapsto v_1$ under the actions of $LSU(n+1)$ and $\mathbb{T}\tilde{\times}LSU(n+1)$ are $LU(n)$ and $\mathbb{T}\tilde{\times}LU(n)$ respectively.  Therefore we may write the loop space $L\mathbb{CP}^n$ as quotients:
\begin{equation} L\mathbb{CP}^n \cong LSU(n+1)/LU(n) \cong \mathbb{T}\tilde{\times}LSU(n+1)/\mathbb{T}\tilde{\times}LU(n). \end{equation}

We will represent points of $L\mathbb{CP}^n$ by column vectors of functions $z^i:S^1\rightarrow \mathbb{C}$ defined up to overall multiplication by functions $f$ from $S^1$ to $\mathbb{C}\backslash\{0\}$:
\begin{equation}
\left[ \begin{array}{c} z^1(\theta) \\ \vdots \\ z^{n+1}(\theta) \end{array} \right] \sim 
\left[ \begin{array}{c} f(\theta) z^1(\theta) \\ \vdots \\ f(\theta) z^{n+1}(\theta) \end{array} \right].
\end{equation}
The functions $z^i$ may not be all zero for any value of $\theta$

An alternative way to represent $L\mathbb{CP}^n$ is as an adjoint orbit.  The Lie algebra of $LSU(n+1)$ consists of maps from $S^1$ to $su(n+1)$, while the Lie algebra of $\mathbb{T}\tilde{\times}LSU(n+1)$ is obtained by adding the generator $\partial_\theta$ of $\mathbb{T}$.  The Lie brackets with $\partial_\theta$ are given by
\begin{equation} [\partial_\theta, Y(\theta) ] = \frac{\partial Y}{\partial\theta}(\theta). \end{equation}
If $b\in su(n+1)$ is the point with stabiliser $U(n)$, then the constant loop $\theta\mapsto b$ has stabiliser $\mathbb{T}\tilde{\times}LU(n)$ under the adjoint action of $\mathbb{T}\tilde{\times}LSU(n+1)$.  Therefore $L\mathbb{CP}^n$ may be identified with the adjoint orbit of $b$.

The space $L\mathbb{CP}^n$ inherits a natural metric from $\mathbb{CP}^n$.  Let $z=z(\theta)$ denote a point in $L\mathbb{CP}^n$.  A tangent vector at $z\in\mathbb{CP}^n$ may be represented by a function $\xi:S^1\rightarrow T\mathbb{CP}^n$, such that $\xi(\theta)\in T_{z(\theta)}\mathbb{CP}^n$.  A natural metric on such tangent vectors is
\begin{equation}
\label{loop metric}
g_z(\xi_1,\xi_2) = \frac{1}{2\pi}\int_{S^1} g_{z(\theta)}(\xi_1(\theta),\xi_2(\theta)) d\theta.
\end{equation}

\begin{figure}
\epsfig{file = 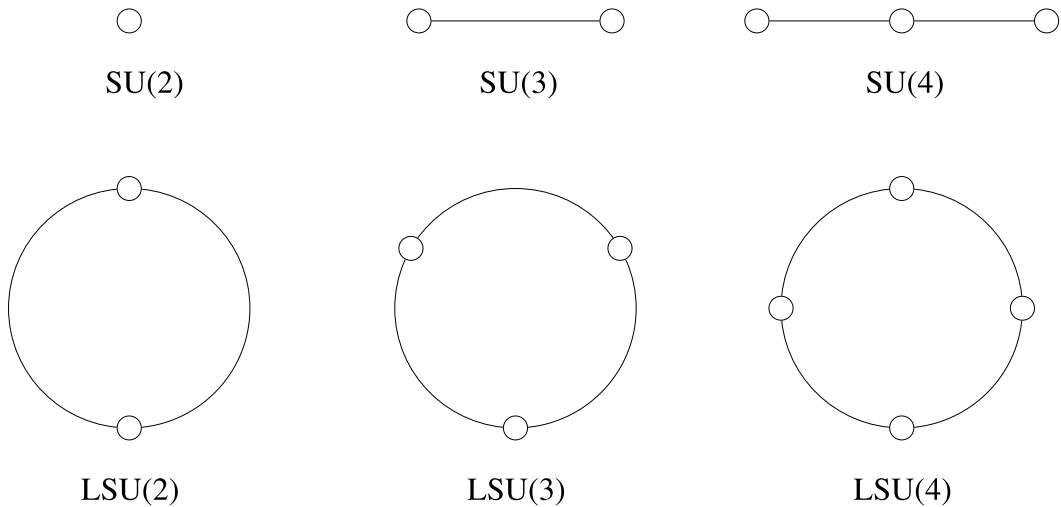, scale = 1}
\caption{The Dynkin diagrams of various Lie groups and their loop groups}
\label{dynkin}
\end{figure}

It will be useful to recall a little of the structure theory for Lie algebras of loop groups.  If $\mathfrak{t}$ is a Cartan subalgebra of $su(n+1)$, then a Cartan subalgebra for the Lie algebra of $\mathbb{T}\tilde{\times}LSU(n+1)$ is obtained by adding the generator $\partial_\theta$ of $\mathbb{T}$.  The roots are denoted $(k,\alpha)$, where $k\in\mathbb{Z}$, $\alpha$ is a root of $su(n+1)$ or zero, and $(k,\alpha)\neq(0,0)$; their action is $(k,\alpha)(X)=\alpha(X)$ for $X\in\mathfrak{t}$, and $(k,\alpha)(\partial_\theta)=ik$.  There are $n+1$ simple roots, given by
\begin{eqnarray}
\underline{\alpha}_i &=& (0,\alpha_i) \mbox{ for }i=1,\dots,n \\
\underline{\alpha}_{n+1} &=& \left( -1,-\sum_{i=1}^n \alpha_i \right).
\end{eqnarray}
The Dynkin diagrams of Lie groups $SU(n+1)$ and the corresponding loop groups are depicted in figure \ref{dynkin}.

\section{A chain as a loop group multi-kink}

We are now ready to elucidate the root structure associated with chains.  We shall show that a $\mathbb{CP}^n$ chain is the same thing as a multi-kink in a sigma model whose target is the loop space $L\mathbb{CP}^n$.  Let $\phi$ be a chain satisfying $\phi(x,y+\beta)=\exp(-\beta X)\phi(x,y)$.  Define a map $\tilde{\phi}:\mathbb{R} \rightarrow L\mathbb{CP}^n$ by
\begin{equation} \tilde{\phi}(x) = \exp(X\theta/\mu_0) \phi(x,\theta/\mu_0). \end{equation}
Making this reparametrisation is a little like changing the gauge of a caloron from the algebraic gauge (where the component of the gauge field in the periodic direction $A_0$ vanishes at infinity and the caloron is periodic up to a gauge transformation) to a gauge where $A_0$ is non-zero at infinity and the caloron is strictly periodic (eg the Polyakov gauge) \cite{vb98}.

%Making this reparametrisation is a little like changing the gauge of a caloron from the algebraic gauge (where the component of the gauge field in the periodic direction $A_0$ vanishes at infinity and the caloron is periodic up to a gauge transformation) to the Polyakov gauge (where $A_0$ is constant and the caloron is strictly periodic).  However, while topological obstructions are encountered in writing a caloron in the Polyakov gauge, our reparametrisation for sigma model chains is valid everywhere.

The Bogomolny equation for $\phi$,
\begin{equation} \left( \frac{\partial}{\partial x} + i\frac{\partial}{\partial y} \right) \phi = 0, \end{equation}
is equivalent to
\begin{equation}
\frac{\partial}{\partial x} \tilde{\phi} = i \left( X - \mu_0 \frac{\partial}{\partial \theta} \right) \tilde{\phi}.
\end{equation}
This is formally identical to the Bogomolny equation for a multi-kink (\ref{dw bog equation}), since $X- \mu_0\partial_\theta$ is an element of the Lie algebra of $SU(n+1) \tilde{\times} \mathbb{T}$.  Similarly, the energy functional for $\phi$ is equal to
\begin{equation} E = \frac{\beta}{2} \int_\mathbb{R} \| \partial_x\tilde{\phi} \|^2 + V(\tilde{\phi}) dx, \end{equation}
where the potential function $V:L\mathbb{CP}^n\rightarrow \mathbb{R}$ is equal to the norm squared of the vector field induced on $L\mathbb{CP}^n$ by $X-\mu_0\partial_\theta$.  Here we are using the metric defined in (\ref{loop metric}).

Identifying chains with $L\mathbb{CP}^n$ kinks explains the constituent structure of chains identified earlier.  Just as the $n$ fundamental kinks of a multi-kink were identified with the $n$ simple roots of $SU(n+1)$, so too are the $(n+1)$ fundamental kinks of a chain identified with the $(n+1)$ simple roots of $SU(n+1) \tilde{\times} \mathbb{T}$.  Notice too that the masses $\nu_i$ of the fundamental kinks are in this case proportional to $\underline{\alpha}_i (X-\mu_0\partial_\theta)$, analagous to (\ref{mass formula}).

\begin{figure}[ht]
\epsfig{file = 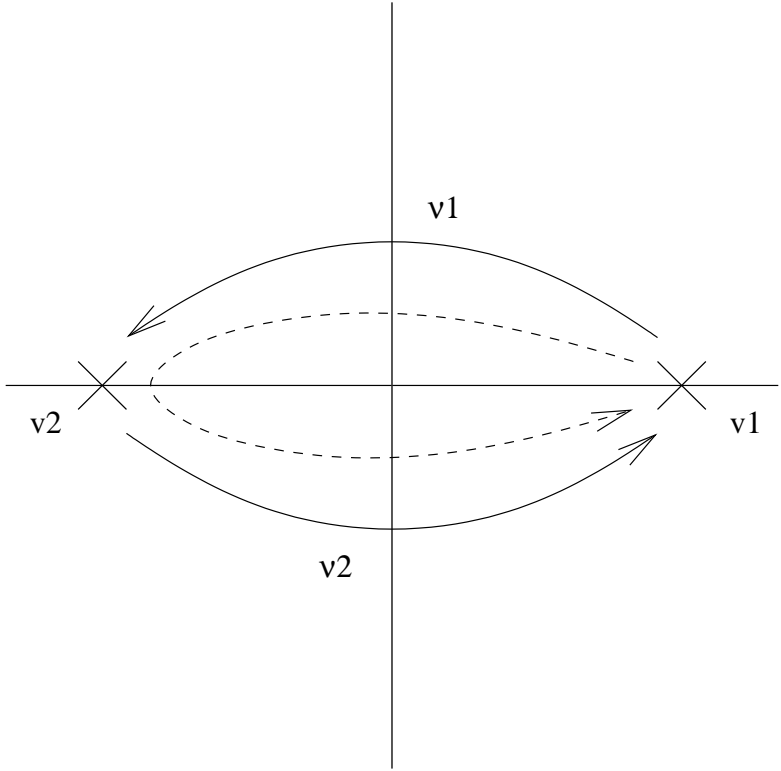, scale = 1}
\caption{The charge 1 $\mathbb{CP}^1$ chain realised as an adjoint orbit in the Lie algebra of the loop groop}
\label{smc fig 2}
\end{figure}

As a concrete example, the Bruckmann chain (\ref{bruckmann chain}) is mapped to
\begin{equation} \tilde{\phi}(x) = \left[ \begin{array}{c} a_{11} \exp((\mu_0-\mu_1)x) e^{i\theta} + a_{10} \exp(-\mu_1x) \\ a_{20} \exp(-\mu_2x) \end{array} \right]. \end{equation}
We have illustrated this object in figure \ref{smc fig 2}.  The horizontal axis represents the Cartan sub-algebra $\mathfrak{t}$ of $su(2)$ and the vertical axis represents the Lie algebra of $\mathbb{T}$.  The orbit of $v_1=t$ is $L\mathbb{CP}^1$.  The dashed arrow represents the chain (\ref{bruckmann chain}), while the two solid arrows indicate the fundamental kinks.  The kink pointing from $v_1$ to $v_2$ has mass $\nu_1$, the kink pointing from $v_2$ to $v_1$ has mass $\nu_2$.

\section{Summary and open problems}

In this article, we have explored the connection between chains and multi-kinks in sigma models.  We were able to construct explicitly all solutions to the appropriate Bogomolny equations, with our choice of boundary conditions.  We showed how the constituent structure of both can be understood using the structure theory of Lie algebras; for the case of chains, the relevant Lie algebra was associated with a loop group.  The constituents provide a simple physical picture of multi-kinks and chains: the number of moduli can be accounted for by assigning a position and a phase modulus to each constituent, while the total mass is equal to the sums of the masses of all of the constituents.

The dimensions of the moduli spaces of chains are completely determined in a simple way by the topological charges and boundary conditions imposed.  We have demonstrated this by constructing all solutions explicitly, but for other solitons (for example instantons, calorons and monopoles) the dimension of the moduli space is known without existence of explicit solutions.  It would be interesting to know whether the dimensions of moduli spaces could be related to the topology more directly, for example by using an index theorem.

We point out here that sigma models have been studied in connection with loop groups before.  Atiyah has demonstrated the existence of a bijection between moduli spaces of $SU(n)$ Yang-Mills instantons and moduli spaces of lumps in sigma models with target $\Omega SU(n+1)=LSU(n+1)/SU(n+1)$ \cite{atiyah84,guest85}, while Charbonneau and Hurtubise have explored a similar correspondence for calorons \cite{charbonneau&hurtubise08}.

We have discovered many similarities between $\mathbb{CP}^n$ sigma model chains and $SU(n+1)$-calorons.  Most notable are the charge formula (\ref{caloron charge}), which resembles the formula for calorons, and the ordering prescription among consituents of a chain, which mirrors an ordering prescripition among the constituent monopoles $SO(2)$-symmetric calorons.  The latter might have an explanation via Atiyah's result, particularly as Atiyah's paper focuses on the case of $SO(2)$ symmetry, and exploits loop groups.  However, we have not succeeded in finding a direct link; the main problem is that Atiyah's target space $\Omega SU(n+1)$ does not coincide with our target space $L\mathbb{CP}^n$.

\subsection*{Acknowledgements}

I would like to thank Richard Ward, Paul Sutcliffe, Martin Speight and Bernd Schroers for comments, suggestions, and discussions.  I am grateful to the authors of \cite{bruckmann09} for informing me of their work prior to publication.  This work was completed while supported by a STFC studentship at Durham University.


\begin{thebibliography}{10}

\bibitem{abraham&townsend2}
E.~R.~C. Abraham and P.~K. Townsend.
\newblock More on Q-kinks: a $(1+1)$-dimensional analog of dyons.
\newblock {\em Phys. Lett.}, B295:225--232, 1992.

\bibitem{abraham&townsend1}
E.~R.~C. Abraham and P.~K. Townsend.
\newblock Q-kinks.
\newblock {\em Phys. Lett.}, B291:85--88, 1992.

\bibitem{aaw02}
Miguel Aguado, M.~Asorey, and A.~Wipf.
\newblock {Nahm transform and moduli spaces of CP(N)-models on the torus}.
\newblock {\em Annals Phys.}, 298:2--23, 2002, hep-th/0107258.

\bibitem{atiyah84}
M.~F. Atiyah.
\newblock Instantons in two and four dimensions.
\newblock {\em Commun. Math. Phys.}, 93:437--451, 1984.

\bibitem{vb02}
Falk Bruckman and Pierre van Baal.
\newblock Multi-caloron solutions.
\newblock {\em Nucl. Phys.}, B645:105--133, 2002, arXiv:hep-th/0209010.

\bibitem{bruckmann08}
Falk Bruckmann.
\newblock Instanton constituents in the O(3) model at finite temperature.
\newblock {\em Phys. Rev. Lett.}, 100:051602, 2008, arXiv:0707.0775.

\bibitem{bruckmann09}
Wieland Brendel, Falk Bruckmann, Lukas Janssen, Andreas Wipf, and Christian Wozar.
\newblock Instanton constituents and fermionic zero modes in twisted CP(n)
  models.
\newblock {\em Phys. Lett.}, B676:116--125, 2009, arXiv:0902.2328.

\bibitem{charbonneau&hurtubise08}
Benoit Charbonneau and Jacque Hurtubise.
\newblock Calorons, Nahm's equations on $S^1$, and bundles over
  $\mathbb{P}^1\times\mathbb{P}^1$.
\newblock {\em Commun. Math. Phys.}, 280:315--249, 2008, arXiv:math/0610804v1.

\bibitem{efinoos06}
Minoru Eto, Toshiaki Fujimori, Youichi Isozumi, Muneto Nitta, Keisuke Ohashih,
  Kazutoshi Ohta, and Norisuke Sakai.
\newblock {Non-Abelian vortices on cylinder: Duality between vortices and
  walls}.
\newblock {\em Phys. Rev.}, D73:085008, 2006, hep-th/0601181.

\bibitem{gm88}
H.~Garland and Michael~K. Murray.
\newblock Kac-Moody monopoles and periodic instantons.
\newblock {\em Commun. Math. Phys.}, 120:335--351, 1988.

\bibitem{gtt01}
Jerome~P. Gauntlett, David Tong, and Paul~K. Townsend.
\newblock Multidomain walls in massive supersymmetric sigma models.
\newblock {\em Phys. Rev.}, D64:025010, 2001, arXiv:hep-th/0012178.

\bibitem{guest85}
M.~A. Guest.
\newblock Instantons and harmonic maps.
\newblock In V.~Kac, editor, {\em Infinite dimensional groups with
  applications}, pages 137--156. Springer-Verlag, 1985.

\bibitem{symcal}
Derek Harland.
\newblock Large scale and large period limits of symmetric calorons.
\newblock {\em J. Math. Phys.}, 48:082905, 2007, arXiv:0704.3695.

\bibitem{vb98}
Thomas~C. Kraan and Pierre van Baal.
\newblock Periodic instantons with non-trivial holonomy.
\newblock {\em Nucl. Phys.}, B533:627--659, 1998, arXiv:hep-th/9805168.

\bibitem{ll}
Kimyeong Lee and Changhai Lu.
\newblock $SU(2)$ calorons and magnetic monopoles.
\newblock {\em Phys. Rev.}, D15:025011, 1998, arXiv:hep-th/9802108.

\bibitem{lee&yi}
Kimyeong Lee and Piljin Yi.
\newblock Monopoles and instantons on partially compactified D-branes.
\newblock {\em Phys. Rev.}, D56:3711--3717, 1997, arXiv:hep-th/9702107.

\bibitem{manton&sutcliffe}
Nicholas Manton and Paul Sutcliffe.
\newblock {\em Topological solitons}.
\newblock Cambridge University Press, 2004.

\bibitem{mottola&wipf89}
Emil Mottola and Andreas Wipf.
\newblock Unsuppressed fermion-number violation at high temperature: an O(3)
  model.
\newblock {\em Phys. Rev.}, D39:588--602, 1989.

\bibitem{nye}
Thomas M.~W. Nye.
\newblock {\em The Geometry of Calorons}.
\newblock PhD thesis, University of Edinburgh, 2001, arXiv:hep-th/0311215.

\bibitem{ps86}
Andrew Pressley and Graeme Segal.
\newblock {\em Loop groups}.
\newblock Oxford University Press, 1986.

\bibitem{romao05}
Nuno~M. Rom\~ao.
\newblock Dynamics of $\mathbb{CP}^1$ lumps on a cylinder.
\newblock {\em J. Geom. Phys.}, 54:42--76, 2005, arXiv:math-ph/0404008.

\bibitem{snippe94}
Jeroen Snippe.
\newblock Tunneling through sphalerons: the O(3) sigma model on a cylinder.
\newblock {\em Phy. Lett.}, B335:395--402, 1994, arXiv:hep-th/9405129.

\bibitem{tong02}
David Tong.
\newblock The moduli space of BPS domain walls.
\newblock {\em Phys. Rev.}, D66:025013, 2002, arXiv:hep-th/0202012.

\bibitem{ward05}
R.~S. Ward.
\newblock Periodic monopoles.
\newblock {\em Phys. Lett.}, B619:177--183, 2005, arXiv:hep-th/0505254.

\bibitem{weinberg80}
Erick~J. Weinberg.
\newblock Fundamental monopoles and multimonopole solutions for arbitrary
  simple gauge groups.
\newblock {\em Nucl. Phys.}, B167:500--524, 1980.

\end{thebibliography}
\end{document}